# Fixing the shadows while moving the gnomon

**Alejandro Gangui**
IAFE/Conicet and Universidad de Buenos Aires, Argentina.

*It is a common practice to fix a vertical gnomon and study the moving shadow cast by it. This shows our local solar time and gives us a hint on the season in which we perform the observation. The moving shadow can also tell us our latitude with high precision. In this paper we propose to exchange the roles, and while keeping the shadows fixed on the ground we will move the gnomon. This lets us understand in a simple way the relevance of the tropical lines of latitude and the behavior of shadows in different locations. We then put these ideas into practice using sticks and threads during a solstice on two sites located on opposite sides of the Tropic of Capricorn.*

PACS: 01.40.-d, 01.40.ek, 95.10.-a

Sundials usually include three lines (called declination lines) drawn on their surfaces (or dial faces). One of these is a straight line which corresponds to the shadow cast by the style (the time-telling edge of the gnomon) during the day on both equinoxes, and is oriented exactly in the east-west direction. The other two lines are curved and are located on opposite sides of the straight one: they also have opposite curvatures, with their concavity pointing away from the middle line. These two lines correspond to the style's shadow during each of the solstices: the one to the north of the middle line is followed by the shadow during the December solstice (and the one to the south, during the June solstice) [1].

All this is easy to see for a vertical gnomon placed exactly on the equator. During northern spring and summer (from approximately March 21 to September 21), as the Sun is located to the north of the celestial equator (and its declination is positive), the shadows of the style will fall to the south of its base [2]. Moreover, the longest shadow at solar noon will be that of June 21, the Northern Hemisphere summer solstice (see Fig.1).

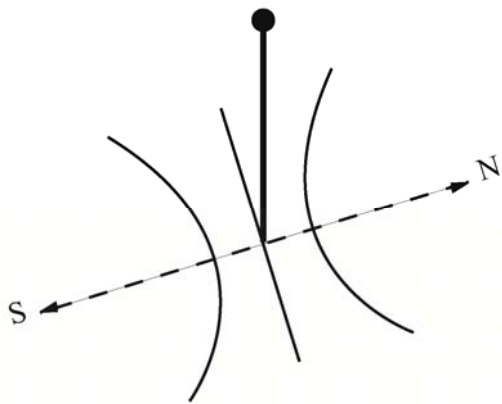

Fig.1: Diagram in perspective showing the north-south (meridian) line and the three declination lines typical of simple sundials. The vertical gnomon corresponds to an observer placed on the equator. It will cast shadows on the floor along curves that fit inside the two extreme ones depicted in this figure. The straight middle line corresponds to the shadows during equinoxes. The curved line to the left corresponds to the shadows during the June 21 solstice.

As the observer moves on from the equator, the three declination lines shift their positions with respect to the base of the gnomon, and also the curvature of the outer two lines gets modified. In Ref. [3] we showed the gradual displacement of the straight middle line for four locations south of the equator: it is shifted towards the south an amount proportional to the increase in the absolute value of the latitude. In general, for small displacements in latitude the difference in the curvature and relative positions of the outer two declination lines is not drastic, but the directions of the shadows at solar noon can surprise many of us.

Both teachers and students sometimes think that the equatorial line is *the* site where astronomical phenomena change drastically, e.g. where the crescent Moon inverts the orientation of its illuminated "face" and, also, where Sun shadows point in opposite directions at noon. This confusion may arise from our habit of speaking in terms of (northern and southern) hemispheres and from our frequent disposition to contrast what we would see were we living in one or the other. In the case of the Sun, its location in the sky (its declination) is actually more important for Astronomy than the terrestrial equator, as we will now see.

**Diurnal Astronomy during a solstice**

During the 2011 December solstice, in collaboration with colleagues from a few South American cities, we carried out a joint shadow-observing activity with the aim to emphasize aspects that were characteristic to each location (like the length of the shadows at noon cast by vertical gnomons) from others that were common to all cities (like the approximately similar concavity of the solstice declination line). This activity was a follow up on the one already reported here [3].

Figure 2 shows two observational settings from different latitudes in Brazil: one in the city of Natal (Rio Grande do Norte) and the other in Caxias do Sul (Rio Grande do Sul). These cities have the (geographical) property of being located on opposite sides of the Tropic of Capricorn.

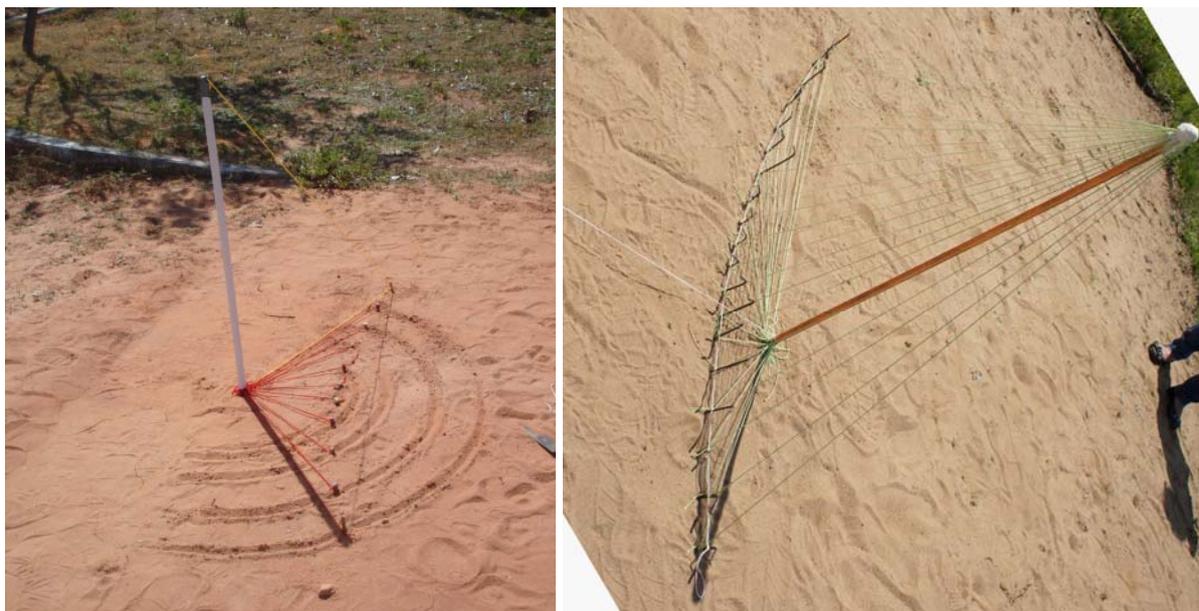

Fig.2: Two observational settings for studying the shadows cast by 1 meter vertical gnomons during the 2011 December solstice. Left image is from Natal (Lat: 5.8°S), while the image to the right is from Caxias do Sul (Lat: 29.2°S). Theoretical values for the angles between the gnomon and the solar noon threads (the shortest ones) are 17.7° = |5.8°-23.5°| for Natal, and 5.7° = (29.2°-23.5°) for Caxias do Sul. Measured values were actually 17° and 7°, respectively. Note that while the concavity of both curves made by the sticks on the ground is approximately the same (the picture on the right was rotated in order to better show this), these declination lines fall on opposite sides of the gnomon: the one in Natal falls to the north of the base of the gnomon; the one in Caixas do Sul, to the south. In both images the north is to the right, as can be inferred by the last shadow cast by the gnomons –and visible in the pictures– which, coming from the southwestern setting Sun, points approximately towards the northeast. Photographs are courtesy of Auta Stella de Medeiros (Natal) and Odilon Giovannini (Caxias).

In both cases, 1 meter vertical gnomons were employed and their shadows were marked on the floor with sticks. Sun rays going from the tip of each gnomon to the sticks were materialized with threads, as shown in the picture from Caxias do Sul (Fig. 2). As we can see, the surprising fact now is to find solar noon shadows falling on opposite sides of the base of the gnomon: the shortest shadow in Caxias do Sul points towards the south; the corresponding one in Natal points towards the north. Clearly, the key latitude is the tropic not the equator: during the December solstice (with the Sun at declination = -23.5°), cities situated north or south of latitude 23.5°S will experience opposite-pointing noon shadows (see Fig. 3). As a corollary of this, those sites located exactly along the Tropic of Capricorn will project no noon shadow at all.

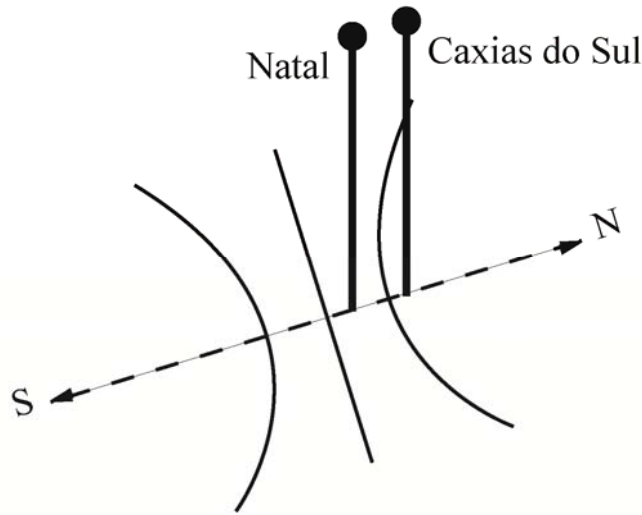

Fig.3: Fixing the shadows and moving the gnomons. This diagram is similar to that of Figure 1, but now for gnomons placed on different latitudes. The one on the equator (Lat: 0°) would be placed on the straight middle line, as required by symmetry. Those located to the south of the equator, for example in Natal (Lat: 5.8°S) and in Caxias do Sul (Lat: 29.2°S), are shown in the diagram. One gnomon located in Buenos Aires (Lat: 34.5°S), would be placed further to the right of these. Hence, shadows observed in Caxias do Sul or in Buenos Aires at noon can never point towards the north. (Although we know that the declination lines are slightly modified if we change the latitude of observers located not very far from the equator, here, for simplicity, we considered them unchanged.)

Of course, a similar observational exercise can be planned for cities located on opposite sides of the Tropic of Cancer, for example San José de Costa Rica (Lat: 9.9°N) and Dallas (Lat: 32.8°N). In this case the observations should be performed during days close to the June solstice. The solar noon shadow projected by the gnomon in Dallas will fall towards the north, while the solar noon shadow projected by the gnomon located in San José will fall towards the south.

In summary, while in general during outdoor diurnal astronomical activities we expect to have a static gnomon fixed to the ground and see the shadows cast by it moving around, here we saw that it is also possible (within an acceptable level of approximation) to keep fixed the declination lines and "move" instead the gnomon. Doing this allows us to predict, without travelling away from our site, what other observers in other locations will actually see. If we stick to latitudes within a few tens of degrees from the equator (to put a number, latitudes below 40°), the declination lines are not altered very much [1]. Displacing the gnomon along the meridian line, as shown here in Figure 3, may let us learn quite simply the relevance of the tropical lines of latitude and may provide us with a way to understand why noon shadows in tropical regions change their orientations during spring and summer in the way they do.

**Acknowledgements**
We thank our colleagues from the project "Oblicuidad de la eclíptica", and specially Professors Auta Stella de Medeiros and Odilon Giovannini, for sharing with us their observations and pictures.

We also thank the referees for their useful comments that helped improve the paper. A.G. acknowledges support from Conicet and from the University of Buenos Aires.

Alejandro Gangui *is staff researcher at the Institute for Astronomy and Space Physics (IAFE) and is professor of physics at the University of Buenos Aires (UBA).*